\documentclass[utf8]{FrontiersinHarvard} 

\usepackage{url,hyperref,microtype,subcaption}
\usepackage[onehalfspacing]{setspace}
\usepackage[english]{babel}
\usepackage{multirow}
\usepackage{booktabs}
\usepackage{algorithm}
\usepackage{algpseudocode}
\usepackage{adjustbox}
\usepackage{makecell}

\def\keyFont{\fontsize{8}{11}\helveticabold }
\def\firstAuthorLast{M\"ullner {et~al.}} 
\def\Authors{Peter M\"ullner\,$^{1,2, *}$, Anna Schreuer\,$^{2}$, Simone Kopeinik\,$^{1}$, Bernhard Wieser\,$^{2}$, and Dominik Kowald\,$^{1,2}$}
\def\Address{$^{1}$ Know Center Research GmbH, Graz, Austria \\
$^{2}$ Graz University of Technology, Graz, Austria}
\def\corrAuthor{Peter M\"ullner}
\def\corrEmail{pmuellner@\{know-center.at, tugraz.at\}}

\newcommand{\rev}[1]{\textcolor{black}{{#1}}}
\newcommand{\revnew}[1]{\textcolor{black}{{#1}}}

\begin{document}
\onecolumn
\firstpage{1}

\title[Multistakeholder Fairness in Tourism]{Multistakeholder Fairness in Tourism: What can Algorithms learn from Tourism Management?} 

\author[\firstAuthorLast ]{\Authors} 
\address{} 
\correspondance{} 

\extraAuth{}

\maketitle

\begin{abstract}

\section{
Algorithmic decision-support systems, i.e., recommender systems, are popular
digital tools that help tourists decide which places and attractions to explore.
However, algorithms often unintentionally direct tourist streams in a way that negatively affects the environment, local communities, or other stakeholders. 
This issue can be partly attributed to the computer science community's limited understanding of the complex relationships and trade-offs among stakeholders \rev{in the real world}.
\rev{In this work, we draw on the practical findings and methods from tourism management to inform research on multistakeholder fairness in algorithmic decision-support.}
\revnew{Leveraging a semi-systematic literature review, we synthesize literature from tourism management as well as literature from computer science.}
Our findings suggest that \rev{tourism management actively tries to identify the specific needs of stakeholders and} utilizes qualitative, inclusive and participatory methods to study fairness from a normative and holistic research perspective.
In contrast, computer science \rev{lacks sufficient understanding of the stakeholder needs and} primarily considers fairness through descriptive factors, such as measureable discrimination, while heavily \rev{relying} on \rev{few} mathematically formalized fairness criteria \rev{that fail to capture the multidimensional nature of fairness in tourism}. 
\revnew{With the results of this work, we aim to illustrate the shortcomings of purely algorithmic research and stress the potential and particular need for future interdisciplinary collaboration.}
We believe such a collaboration is a fundamental and necessary step to enhance algorithmic decision-support systems towards understanding and supporting true multistakeholder fairness in tourism.

}

\tiny
 \keyFont{ \section{Keywords:} Tourism, Recommender Systems, Decision-Support, Interdisciplinary Research, Multistakeholder Fairness} 

 \noindent \keyFont{This is a preprint of a paper accepted by Frontiers in Big Data, Section recommender systems, research topic on ``Guiding the Journey: Innovative Recommender Systems for Personalized Tourism, Travel, and Hospitality Experiences'', DOI: 10.3389/fdata.2025.1632766}
\end{abstract}

\section{Introduction}

Tourism contributes significantly to economic growth \citep{wijesekara2022tourism,li2018tourism}. 
Depending on the specific region, the economic benefit for the local community can be substantial.
Yet, not only between, but also within a given region, the distribution of benefits among different groups of society may vary. 
Who stands to profit, and who may not gain anything, or may even face negative implications that come along with tourist activities? 
Revealing the ways in which the benefits of tourism are distributed throughout society is a complex task~\citep{dangi2021augmenting}. 
It is especially important to consider those who are not directly engaged in business transactions of the tourist industry~\citep{banerjee_fairness_2023}. 
Local residents may be affected by rising house prices, the effects of tourist activities on the environment may be substantial, and everyday life in general may be impacted in undesirable ways~\citep{van2018platform}. 
\rev{Especially for destinations in the Global South, researchers have highlighted that local communities do not appropriately benefit from tourism and have explored how sustainable tourism can remedy such injustices~\citep{brune2022sustainable,rastegar2022injustices}}

The problems mentioned above increase with the concentration of large numbers of tourists in popular destinations. 
Overtourism has been recognized to be a key challenge in the industry and technical solutions that help redistribute tourists more evenly are highly sought after~\citep{banerjee_fairness_2023}. 
In order to tackle these challenges, fairness has become a central conceptual reference point and a main requirement for trustworthy AI~\citep{kowald2024establishing}.  
Fairness draws attention to possible inequalities among users of a service or product. 
Such inequalities may bear on individuals, but they may also play out between entire groups of society. 
Certain attributes such as gender, age, and ethnicity are considered highly sensitive and no discrimination should be made based on these attributes. 
More recent studies~\citep{abdollahpouri2019multi,burke2022multi,sonboli2022multisided} have expanded the scope of the fairness discussion to a multistakeholder perspective in order to account for the diversity of needs and interests between the various social actors and groups.
It is especially challenging to implement heterogeneous fairness dimensions, and it is even more challenging if such fairness dimensions are not easily quantifiable or difficult to weigh against each other.

\noindent \textbf{Decision-Support for Multistakeholder Fairness in Tourism.} 
Today, tourists increasingly rely on algorithmic decision-support, i.e., recommender systems, to discover points-of-interest, e.g., destinations, accommodations, or attractions, that match their preferences~\citep{borras2014intelligent,ricci2022recommender,sanchez2022point}. 
However, the influence of such algorithms extends beyond individual tourists as they can redirect tourism flows and impact local communities, businesses, and the environment~\citep{balakrishnan_multistakeholder_2021}. 
While these systems are designed to optimize end-user satisfaction, they struggle to ensure multistakeholder fairness due to a lack of understanding of the complex trade-offs between various stakeholders~\citep{sonboli2022multisided,atzenhofer2024value,atzenhofer2025multistakeholder,burke2024multistakeholder,burke2025centering}. 
Failing to consider stakeholder needs contributes to issues such as overtourism, environmental pollution, unaffordable housing for residents, or unfair distribution of the economic benefit across stakeholders~\citep{banerjee2025modeling}.
A key limitation is the way fairness is incorporated into algorithmic decision-support systems:
The computer science community primarily frames fairness through quantifiable criteria, often focusing on \revnew{algorithmic} approaches~\citep{deldjoo2024fairness} that mitigate discrimination or bias, e.g., popularity bias~\citep{kowald2020unfairness,mullner2023reuseknn,kowald2022popularity}, which favors popular destinations over others~\citep{rahmani_unfairness_2022,forster2025}.
This results in algorithmic decision-support systems that fail to implement the complex fairness criteria that are required for the numerous stakeholders in tourism.
In contrast, the tourism management community tends to consider fairness \rev{as} a complex, multidimensional issue that involves diverse stakeholders with competing interests. 
The lack of interdisciplinary collaboration results in algorithmic decision-support systems that struggle to adequately incorporate multistakeholder fairness, as they fail to account for the complex relationships between tourists, local residents, governments, and businesses. Additionally, this lack of interdisciplinary collaboration hinders a concrete operationalization of fairness goals into metrics and algorithms \citep{smith2023scoping}, which can be integrated into decision support and recommendation frameworks (e.g., \citep{lacic2014towards,tourani2024capri}). 

\noindent \textbf{Research Questions and Findings.} \revnew{To better connect the algorithm-focused computer science research with the findings from tourism management, we review literature from two domains and investigate how they compare.}
Specifically, we address \emph{RQ1: How do \revnew{tourism management and computer science} differ in their fairness definitions?}
 and \emph{RQ2: How can algorithm design benefit from \rev{the research body on tourism management?}} 
Through a semi-systematic literature review, we analyze existing research, highlighting differences and similarities in their fairness \rev{definitions}. 
Our findings suggest that the \rev{tourism management} community qualitatively studies multistakeholder fairness, incorporating diverse stakeholders with competing interests. 
In contrast, the computer science community focuses on quantitative fairness criteria that can be included in algorithmic decision-support systems, but covers only a narrow perspective on fairness.
Overall, we hope that the publication at hand underscores that collaboration among the two research communities is crucially needed for developing algorithmic decision-support systems that can successfully balance the needs of numerous stakeholders in the ever-growing tourism industry. 

\noindent \textbf{Structure of this Paper.} Our work is structured as follows: 
Section~\ref{sec:method} details the review methodology used to identify and filter relevant literature. 
Section~\ref{sec:ss_perspective} synthesizes findings from the tourism management perspective, whereas Section~\ref{sec:cs_perspective} presents the algorithmic perspective from computer science. 
Finally, Section~\ref{sec:findings} discusses the key differences and intersections between the two domains, and identifies open research directions.

\section{Review Methodology}
\label{sec:method}
\begin{figure}
    \centering
    \includegraphics[width=0.8\linewidth]{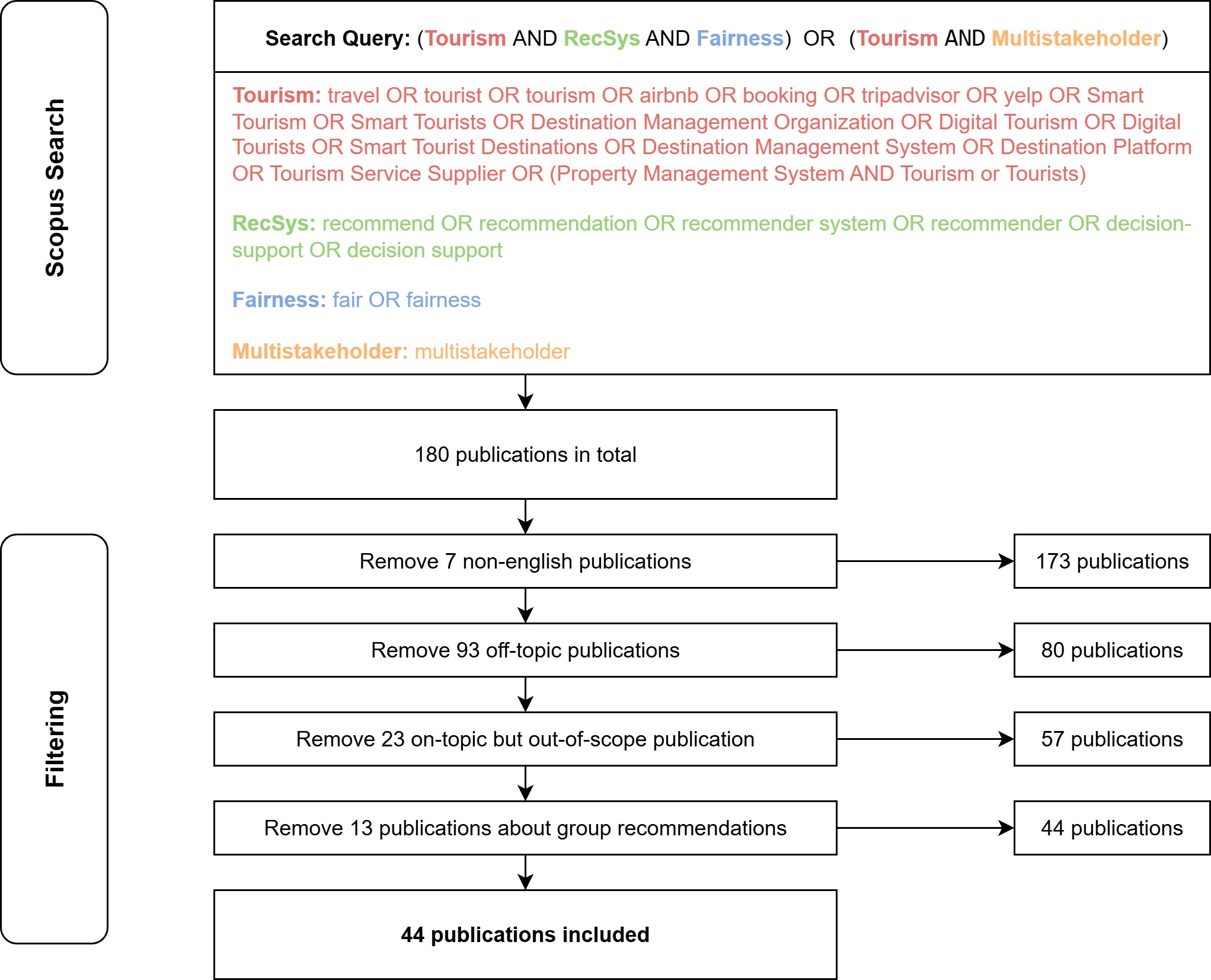}
    \caption{\revnew{Our methodology for identifying and filtering relevant publications. Overall, we select 44 publications for inclusion in this review article.}}
    \label{fig:method}
\end{figure}

For this review article, we performed a Scopus\footnote{https://www.scopus.com/} search to identify publications related to multistakeholder fairness in tourism from both the \rev{tourism management} and the computer science domain.
Through a preliminary manual search, we compiled a set of suitable search terms to design a query that covers the relevant concepts from both research domains.  
The final search query includes search terms related to fairness and decision-support in tourism, as well as recommender systems and fairness among multiple stakeholders (see Figure~\ref{fig:method}).
This search query delivered 180 publications on which we performed several post-filtering steps:
First, we removed any publications not available in English and those clearly off-topic.
This reduced the number of publications from 180 to 80.


Next, we investigated this 80 publications in more detail to ensure their relevance to our specific research focus. 
In this stage, we excluded publications that, although related to the topic itself, did not align with the scope of this study, e.g., studies focusing on marketing strategies or research centered on group recommendation systems.

After applying these post-filtering steps, a final set of 44 publications remained (see Table~\ref{tab:all_papers}), which constitutes the literature reviewed and discussed in this article.
In addition, to ensure reproducibility~\citep{semmelrock2025reproducibility} of our research, we publish the set of publications after each individual filtering step in our GitHub repository\footnote{https://github.com/pmuellner/FairRecSys}.

\begin{table}[!t]
    \centering
    \begin{tabular}{l c c c c}
    \toprule
    & \multicolumn{2}{c}{Research Domain} \\ \cmidrule(lr){2-3}
    Reference & \rev{Tourism Management} & Computer Science \\ \midrule
    \cite{banerjee2025modeling} &  & $\bullet$ \\
    \cite{rodriguez2025prosocial} & $\bullet$ & \\
    \cite{samal2025strengths} & $\bullet$ & \\
    \cite{banerjee_green_2024} & & $\bullet$ \\
    \cite{hasayotin_empowerment_2024} & $\bullet$ & \\
    \cite{khaili2024multi} & & $\bullet$ \\
    \cite{khatri2024tourism} & $\bullet$ & \\
    \cite{merinov2024positive} & & $\bullet$ \\
    \cite{pereira-moliner_conducting_2024} & $\bullet$ & \\
    \cite{romeo2024mountain} & $\bullet$ & \\
    \cite{sarhan2024tourism} & $\bullet$ & \\
    \cite{solano2024enhancing} & & $\bullet$ \\
    \cite{yeager2024process} & $\bullet$ & \\
    \cite{banerjee_review_2023} & & $\bullet$ \\
    \cite{banerjee_fairness_2023} & & $\bullet$ \\
    \cite{banik_understanding_2023} & & $\bullet$ \\
    \cite{chan_sustainable_2023} & $\bullet$ & \\
    \cite{merinov_sustainability-oriented_2023} & & $\bullet$ \\
    \cite{patro2023algorithmic} & & $\bullet$ \\
    \cite{trang_mainstreaming_2023} & $\bullet$ & \\
    \cite{yudistira2023towards} & $\bullet$ & \\
    \cite{blanco-cerradelo_sustainable_2022} & $\bullet$ & \\
    \cite{majdak_pre-emptively_2022} & $\bullet$ & \\
    \cite{merinov_sustainability_2022} & & $\bullet$ \\
    \cite{rahmani_role_2022} & & $\bullet$ \\
    \cite{rahmani_unfairness_2022} & & $\bullet$ \\
    \cite{rahmani_exploring_2022} & & $\bullet$ \\
    \cite{sitikarn_coffee_2022} & $\bullet$ & \\
    \cite{balakrishnan_multistakeholder_2021} & & $\bullet$ \\
    \cite{biswas2021toward} & & $\bullet$ \\
    \cite{sanchez_effects_2021} & & $\bullet$ \\
    \cite{shen_sar-net_2021} & & $\bullet$ \\
    \cite{sigala2021sharing} & $\bullet$ & \\
    \cite{banerjee_analyzing_2020} & & $\bullet$ \\
    \cite{ikhtiagung_new_2020} & $\bullet$ & \\
    \cite{su202020} & $\bullet$ & \\
    \cite{wu_fast_2020} & & $\bullet$ \\
    \cite{higgins-desbiolles_sustainable_2018} & $\bullet$ & \\
    \cite{mudzengi2018challenges} & $\bullet$ & \\
    \cite{ariffin_sustainable_2017} & $\bullet$ & \\
    \cite{haddock-fraser_multistakeholder_2012} & $\bullet$ & \\
    \cite{plummer_managing_2009} & $\bullet$ & \\
    \cite{adams_who_2003} & $\bullet$ & \\
    \cite{jamal2002beyond} & $\bullet$ & \\ \midrule
    44 Publications &  24 & 20 \\
    \bottomrule
    \end{tabular}
    \caption{\revnew{The 44 reviewed publications sorted by publication year. We mark whether the publication originates from the \rev{tourism management} or the computer science domain. 
    Overall, we investigate 24 publications from \rev{tourism management} and 20 algorithm-focused publications from computer science.}}
    \label{tab:all_papers}
\end{table}

\section{Tourism Management Perspective}
\label{sec:ss_perspective}
\noindent \textbf{Collaborative Decision-Making across Stakeholders.} 
\label{ss:31}
In tourism management, effective multistakeholder governance is crucial for balancing the competing interests of various groups, such as local communities, businesses, and government entities. 
\citet{ikhtiagung_new_2020} highlight the importance of stakeholder collaboration in ecologically sensitive areas, emphasizing the need for local businesses and communities to \rev{actively} participate in tourism planning to improve quality of life and community satisfaction.
In this respect, \citet{jamal2002beyond} investigate collaborative planning processes in protected areas, such as national parks, where stakeholders must be willing to compromise to align their diverse objectives.
A similar approach is outlined by \citet{plummer_managing_2009}, who emphasize the benefits of adaptive co-management in tourism, arguing that such frameworks allow stakeholders to contribute meaningfully to decision-making, even when decision-makers cannot fully comprehend all trade-offs due to bounded rationality. 
Moreover, \citet{sarhan2024tourism} explore how sustainable tourism can be achieved through well-constructed multistakeholder partnerships, noting that such partnerships should be evaluated through a framework that includes ecological, economic, social, and cultural impacts.
\citet{haddock-fraser_multistakeholder_2012} \rev{study dive tourism in Malaysia and} highlight how different stakeholders \rev{along the value chain} perceive the impacts of tourism, with some focusing on economic gains while neglecting environmental sustainability. 
Related research also highlights the importance of responsible research, advocating for multistakeholder involvement to address societal issues beyond academic goals \citep{pereira-moliner_conducting_2024}.

\noindent \textbf{Community Empowerment and Local Benefits.} 
\label{ss:32}
Empowering local communities is essential for ensuring that tourism provides tangible benefits to the people who live in tourist areas: \citet{mudzengi2018challenges} argue that although tourism can offer benefits such as employment and infrastructure improvements, local communities often struggle to capitalize on these opportunities due to limited entrepreneurial skills, lack of capital, and inadequate awareness.
In contrast, \citet{samal2025strengths} show how ecotourism in India has enhanced local livelihoods, with a focus on multistakeholder participation and the development of infrastructure and capacity-building programs.
Furthermore, research explores how small and medium-sized enterprises can diversify local economies and reduce reliance on tourism through digital innovation and government support \citep{hasayotin_empowerment_2024}. 
The work by \citet{sitikarn_coffee_2022} also sheds light on the value of community-based tourism, using the example of coffee production in Thailand, where local involvement and innovation have led to the creation of diverse tourism offerings that benefit the community both economically and socially.
\citet{chan_sustainable_2023} discuss how sustainable practices in rural tourism are driven by local community collaboration, government support, and a focus on environmental and social development.
Similarly, \citet{yudistira2023towards} use user studies and \rev{predictive modeling} to \rev{estimate} regional development, suggesting that local involvement and long-term benefits should be prioritized. 
Research also emphasizes the role of community engagement in tourism development, noting that mapping community assets can help identify resources that support sustainable social, economic, and environmental development \citep{yeager2024process}.
\revnew{\cite{romeo2024mountain} present an initiative, which emphasizes the critical role of mountain regions for sustainable tourism development and supports value chain development to ensure that benefits flow equitably to all regional stakeholders.}
\revnew{\cite{sigala2021sharing} reviews literature on sharing and platform economy in tourism, emphasizing the roles of key stakeholders: platforms, providers, users, and the environment. Digital sharing platforms like Airbnb and Uber are rich financial opportunities for locals, but can lead to complex socioeconomic, or ethical issues, e.g., gentrification. }
Finally, \citet{adams_who_2003} explore the challenges associated with revenue distribution in tourism, highlighting the competing interests of local, national, and international stakeholders, and pointing out the difficulties in ensuring that local communities receive fair compensation for their participation in tourism.

\noindent \textbf{Sustainable and Ecotourism Development.} 
\label{ss:33}
Sustainable tourism development seeks to balance the needs of tourists with the preservation of the environment as outlined by \citet{higgins-desbiolles_sustainable_2018}. 
The authors argue that the tourism industry’s current growth-oriented mindset is incompatible with true sustainability, suggesting that tourism must \rev{respect} ecological and social limits.
The research by \citet{trang_mainstreaming_2023} explores how ecotourism can be promoted effectively in Vietnam, emphasizing the role of so-called Destination Management Organizations in managing local challenges and improving tourist satisfaction. 
Moreover, \citet{majdak_pre-emptively_2022} propose distributing tourism across rural areas to reduce overtourism in popular destinations, arguing that this strategy brings economic benefits to less-visited areas \rev{and leads to a fairer distribution of tourists}.
Research also shows that when tourists \rev{believe that all stakeholders are treated in a fair way}, they are more likely to engage in prosocial behaviors, such as positive word-of-mouth, which can further support sustainable tourism practices \citep{rodriguez2025prosocial}. 
In the domain of thermal tourism, \citet{blanco-cerradelo_sustainable_2022} identify several factors that impact the sustainability of tourism across economic, environmental, and social dimensions, underscoring the need for a \rev{holistic} approach to tourism management.
\revnew{\cite{su202020} highlight that there is a trade-off between protecting cultural sites and exploiting touristic value. Plus, stakeholders, such as heritage conservation groups, tourist agencies, or local business and residents, differ in their objectives and values, which makes it hard to find a common ground in decision-making processes.}
\revnew{Finally, \cite{khatri2024tourism} use questionnaires to investigate barriers and challenges of tourism stakeholders, such as hoteliers or tour operators. The key challenges include the protection of the environment and cultural assets, and the incorporation of uniform socio‑cultural, and techno‑environmental constraints across all stakeholders. 
}

\section{Computer Science Perspective}

\label{sec:cs_perspective}

\noindent \textbf{Bias and Multistakeholder Fairness in Recommender Systems.} 
\label{ss:41}
\citet{biswas2021toward} emphasize that most recommender systems are optimized for user satisfaction, which can result in unfair exposure for items, such as points-of-interest (POIs). 
Conversely, systems optimized for item exposure may lead to unfair user experiences. 
To address this two-sided fairness problem, they propose an algorithm that guarantees a minimum exposure level for items while distributing the loss in recommendation quality evenly across users, which ensures envy-freeness~\citep{arnsperger1994envy,burke2022multi}. 
Similarly, \citet{wu_fast_2020} tackle fairness under capacity constraints, proposing a reranking strategy that modifies recommendation lists over multiple rounds. 
This ensures fairness while accounting for the limited capacity of venues like restaurants. 
\citet{rahmani_unfairness_2022} show that POI \rev{recommender systems} often suffer from popularity bias, favoring active users and popular destinations. 
While many models perform well in accuracy, they fail to provide balanced fairness across users and items. 
A related bias is temporal: \citet{rahmani_exploring_2022} demonstrate that users seeking leisure-time recommendations receive preferential treatment over those searching during work hours, despite equal interaction histories. 
Moreover, research identifies position bias in POI \rev{recommender systems}, where nearby but lower-quality venues are ranked higher than more distant yet highly rated establishments, negatively impacting the exposure of deserving businesses \citep{banerjee_analyzing_2020}.
Additionally, \citet{merinov2024positive} simulate tourists' limited knowledge of tourist sites and demonstrate that personalization can be balanced with sustainability to some degree. 
They show that a standard recommender system can promote less-visited locations while maintaining user satisfaction. 
In the context of personalized, scenario-specific travel recommendations, related research introduces SAR-Net \citep{shen_sar-net_2021}. 
This model uses attention mechanisms and scenario-specific transformations to improve fairness and accuracy, while its fairness coefficient corrects exposure bias across different user scenarios.
\citet{rahmani_role_2022} also tackle fairness and overtourism with a recommender system optimized for item and user fairness. 
They employ metrics such as Coverage~\citep{silveira2019good}, Novelty~\citep{gunawardana2012evaluating}, Generalized Cross-Entropy~\citep{deldjoo2021flexible}, and Mean Absolute Deviation~\citep{melchiorre2021investigating} to evaluate fairness and identify popularity and bias.
\citet{balakrishnan_multistakeholder_2021} examine how tourism recommender systems can incorporate the needs of tourists, locals, and service providers. 
They discuss inter- and intra-stakeholder dynamics and show, through user studies, that users are sensitive to other stakeholders' needs. 
External influences such as legislation or seasonal effects are also considered. 
\citet{banerjee_review_2023} argue for balancing the sometimes competing interests of different stakeholders, highlighting the importance of fairness not only for users and providers but also for the broader society and environment.
\revnew{\cite{patro2023algorithmic} examine fairness in recommendation systems on online platforms that include multiple stakeholders. Mainly, they find that most fairness research has focused on settings with two stakeholders, and that improving individual fairness for one stakeholder group often reduces utility for another stakeholder group. This highlights the need for further exploration of multistakeholder fairness involving three or more groups, where the trade-offs become more complex and less well understood.}




\noindent \textbf{Sustainability-oriented Recommendations.} 
\label{ss:42}
Several studies aim to align tourism recommendations with sustainability goals. \citet{banerjee2025modeling} propose a \rev{recommender system} that incorporates CO\textsubscript{2} emissions, destination popularity, and seasonality into travel recommendations. 
Their user study confirms that users are willing to \rev{trade} utility for sustainability. 
\citet{merinov_sustainability_2022} design itineraries that avoid overcrowded POIs while preserving user satisfaction by estimating both utility and environmental impact. 
Similarly, recent research presents a recommender system that promotes societal fairness by recommending environmentally friendly and seasonally balanced destinations~\citep{banerjee_green_2024}.
\citet{banik_understanding_2023} explore user perceptions of sustainability in Venice and find that offering one sustainable alternative per unsustainable choice increases user satisfaction, especially when accompanied by explanations. 
Furthermore, \citet{banerjee_fairness_2023} provide a roadmap for incorporating societal fairness into tourism \rev{recommender systems} by balancing stakeholder concerns, including those of local residents and the environment. 
Related research also explores how to reduce popularity bias and crowding through time-sensitive, stakeholder-aware recommendation techniques \citep{merinov_sustainability-oriented_2023}. 
A broad overview is offered by \citet{banerjee_review_2023}, who review fairness in tourism \rev{recommender systems} from the perspectives of users, providers, items, and society, identifying a gap in \rev{fairness} research \rev{pertaining to issues such as society and sustainability}.


\noindent \textbf{Addressing Data Sparsity for Emerging Destinations.} 
\label{ss:43}
\rev{Data sparsity is especially prevalent for novel, emerging tourist destinations and therefore, recommender systems often fail to recommend such destinations.}
\rev{To resolve this issue, }\citet{solano2024enhancing} develop a rule-based system that uses hierarchical criteria like distance and cuisine to recommend restaurants, especially in emerging tourist regions where data is limited. 
Similarly, \citet{sanchez_effects_2021} suggest merging datasets from different cities and distinguishing between tourists and locals to improve POI recommendations. 
However, their findings indicate that tourists often benefit more from such systems than locals, which can be perceived as unfair. 
\citet{khaili2024multi} propose a multi-funnel architecture to address the cold-start issue~\citep{wei2021contrastive,lacic2015tackling} for newly listed items in travel platforms. 
By separating and then merging cold-item rankings with regular items, their system improves platform diversity and long-term partner retention with minimal \rev{immediate performance losses}. 


\section{Findings and Outlook}
\label{sec:findings}
\begin{table}[]
    \centering
    \begin{adjustbox}{width=\linewidth}
    \color{black}\begin{tabular}{l l l}
    \toprule
    Fairness Definition & Examples of Tourism Management Perspectives & Examples of Algorithmic Perspectives \\ \midrule \midrule
    
    Regional benefits 
    & Higher quality of life~\citep{ikhtiagung_new_2020} & Recommendation coverage \& diversity~\citep{rahmani_role_2022} \\
    & Higher employment rate~\citep{mudzengi2018challenges} & Ensure exposure for regional businesses~\citep{biswas2021toward}\\
    & Long-term regional development~\citep{yudistira2023towards} & \\\midrule
    
    Inclusive decision-making
    & Touristic co-management~\citep{plummer_managing_2009} & Aggregated fairness~\citep{balakrishnan_multistakeholder_2021} \\
    & Establishing partnerships~\citep{sarhan2024tourism} &  \\
    & Align stakeholder objectives~\citep{jamal2002beyond} &  \\
    \midrule
    
    Environmental health & Respecting ecologic limits~\citep{higgins-desbiolles_sustainable_2018} & Balancing fairness with CO\textsubscript{2} emissions~\citep{banerjee2025modeling} \\
    & Redirection of tourists~\citep{majdak_pre-emptively_2022} & Distribute POI popularity~\citep{merinov_sustainability_2022}\\
    & Sustainable behavior~\citep{rodriguez2025prosocial} &   \\ \midrule
    
    Fairness for tourists & Balance tourist needs and sustainability~\citep{higgins-desbiolles_sustainable_2018} & Group fairness based on travel type~\citep{rahmani_exploring_2022} \\ & & Group fairness of tourists vs. locals~\citep{khaili2024multi} \\ & & Group fairness based on activity level~\citep{rahmani_unfairness_2022} \\ \midrule
    
    Exposure for businesses & Increase opportunities along the value chain~\citep{sitikarn_coffee_2022}  & Promoting unpopular POIs~\citep{banerjee_analyzing_2020} \\ & & Addressing data sparsity of new POIs~\citep{solano2024enhancing} \\ & & Fairness under capacity constraints~\citep{wu_fast_2020} \\ \bottomrule
    \end{tabular}
    
    \end{adjustbox}
    
    \caption{
    \rev{Summary of overarching fairness definitions and examples of how fairness is conceptualized in both, tourism management and algorithm-focused computer science literature.}}
    \label{tab:fairness_criterias} 
\end{table}

Overall, we find that fairness is conceptualized differently across the two research domains, i.e., \rev{tourism management} and computer science \rev{(see Table~\ref{tab:fairness_criterias} for an overview).}
\rev{Tourism management tries to actively resolve conflicting interests between stakeholders by acknowledging that fairness is a complex and multidimensional issue that spans social, economic, and environmental dimensions.
It examines the competing interests and power asymmetries between different stakeholder groups and how these play out in specific local contexts.}
In contrast, computer science\rev{'s concept of fairness is strongly biased by whether it can be integrated into algorithms via quantitative metrics. 
Such metrics quantify fairness through comparing the outcomes for different stakeholder groups, which is often a stark oversimplification of the complex interaction patterns and trade-offs between stakeholders.}
\rev{Tourism management emphasizes fairness definitions such as improved regional development, inclusive decision-making, environmental health, and support for local businesses. In contrast, the algorithmic perspective promotes recommendation diversity to support regional equity, uses aggregated fairness to represent multiple stakeholder interests, balances fairness with environmental impact (e.g., CO\textsubscript{2} emissions), ensures group fairness among tourist types, and improves exposure for less popular or new businesses.}
\rev{Moreover, tourism management} typically adopts participatory approaches, actively engaging local communities and affected groups, implementing a bottom-up approach, to define what fairness means in specific contexts. 
\rev{In doing so, this body of work not only addresses distributive fairness (fair outcome), but also strives to promote procedural fairness (fair decision making process).}
Conversely, computer science tends to follow a top\rev{-}down approach, applying generalized fairness models, often developed in isolation from real-world stakeholders (e.g., local communities, small businesses, or the environment).
This failure to properly understand context may result in unintended consequences such as unequal economic benefit distribution, high environmental costs of tourism, or a lack of sustainable and inclusive regional development.
Furthermore, while algorithmic recommender systems have the potential to influence tourist behavior in positive ways, few are designed with sustainability goals in mind, such as promoting low-emission travel, alleviating overcrowding in tourist hotspots, or elevating lesser-known destinations. 
A key finding of our work is that the \rev{practical insights from tourism management} have large potential to enhance algorithmic design by offering a broader, more context-aware understanding of fairness that accounts for various stakeholder's needs~\citep{smith2025pragmatic}. 
Specifically, we have identified three benefits of strengthening interdisciplinary collaboration between \revnew{research in tourism management} and \rev{algorithm-focused research in} computer science:

\noindent \textbf{Providing a Holistic Understanding of Fairness.} 
Recommender systems often reduce fairness to dimensions for which quantifiable metrics are available (e.g., bias mitigation or non-discrimination). 
This overlooks the numerous remaining dimensions of fairness.
\rev{Tourism management} emphasizes fairness as a multidimensional and context-dependent concept, considering the needs of diverse stakeholders, providing a more holistic understanding of fairness than computer science. 
This \revnew{management} perspective can guide algorithm designers to redefine or expand fairness definitions beyond pure technical and quantifiable metrics.

\noindent \textbf{Stakeholder Mapping to Inform Algorithmic Fairness.} 
Recommender systems tend to focus narrowly on users and providers, often neglecting less visible stakeholders like local communities or the environment.
Therefore, they struggle to integrate environmental and societal goals (e.g., reducing overtourism, supporting local businesses) into the recommendation and evaluation process.
The \rev{tourism management} literature offers detailed information on how to perform stakeholder mapping to investigate and \rev{manage} the different impacts on various stakeholders. 
This can inform the design of recommender systems that better reflect the complex trade-offs in the real world.


\noindent \textbf{Guiding Participatory and Inclusive Design.}
The design of recommendation algorithms often excludes those affected by its outcomes from the development process, leading to fairness mismatches.
Plus, they lack transparency and often fail to explain fairness trade-offs in ways that are understandable to users and stakeholders.
\rev{Tourism management research mainly relies on} inclusive, participatory methods (e.g., stakeholder workshops, community-based planning) that can be integrated into the recommender design lifecycle, ensuring that fairness definitions are co-created and context-sensitive.


\revnew{A key challenge is that qualitative fairness goals from tourism management, such as addressing overtourism, cannot be directly optimized as computational metrics for decision support and recommender systems. Instead, their role is to act as guiding principles that inform the selection and design of measurable proxy metrics. This process, known as operationalization \citep{smith2023scoping}, translates an abstract fairness concept into a concrete indicator that can be implemented. For example, the qualitative goal of addressing overtourism (see Section \ref{sec:ss_perspective}) can be operationalized through popularity bias mitigation techniques (see Section \ref{sec:cs_perspective}).  Thus, through this translation, the abstract fairness principles of tourism management are transformed into computable targets and metrics, enabling the system to be aligned with broader goals of sustainability and equity that go beyond predictive accuracy and bias mitigation. Another way to translate fairness goals into metrics is to build on established tourism ecolabels and further certification schemes for responsible tourism (cf.~\citet{jog2024stakeholder}). Recommendation algorithms can thus be designed to prioritize certified businesses or destinations, thus building on pre-existing translations of qualitative goals concerning fair tourism into explicit, verifiable criteria.}

\section{Conclusion}
In this work, we performed a semi-systematic review of multistakeholder fairness in tourism based on 44 publications from the \rev{tourism management} and computer science domains.
This \rev{reveals a substantial} gap between how \revnew{research in tourism management and algorithm-focused research from computer science}, conceptualize \rev{multistakeholder} fairness in tourism. 
While \rev{tourism management} emphasizes a context-aware, stakeholder-\rev{sensitive}, and holistic understanding of fairness, computer science tends to reduce fairness to quantifiable metrics.
\rev{This often} overlooks the complex interactions between diverse stakeholders, including local communities, small businesses, and the environment. 
\rev{Insights} from \rev{tourism management can} enrich algorithmic fairness by contributing valuable insights into complex interaction dynamics and with this, broadening the computer science\rev{'s} perspective on fairness. 
This allows recommender systems to evolve from purely technical tools to instruments that genuinely support fair and sustainable tourism.
Overall, we believe that stronger collaboration between both communities is \rev{essential to establish responsible and fair algorithmic decision-support for tourism.}

\section*{Acknowledgments}
The work received funding from the TU Graz Open Access Publishing Fund, from the FFG COMET program, and from Zukunftsfonds Steiermark.

\bibliographystyle{Frontiers-Harvard} 
\bibliography{bibliography}


\end{document}